\documentclass[runningheads]{svmult}

\usepackage{makeidx}   % allows index generation
\usepackage{graphicx}  % standard LaTeX graphics tool
\usepackage{subeqnar}  % subnumbers individual equations
\usepackage{multicol}  % used for the two-column index
\usepackage{physprbb}  % modified textarea for proceedings,
\makeindex             % used for the subject index
\usepackage{amssymb}
\newcommand{\beq}{\begin{equation}}
\newcommand{\eeq}{\end{equation}}
\newcommand{\barr}{\begin{eqnarray}}
\newcommand{\earr}{\end{eqnarray}}
\newcommand{\andy}[1]{ }
\def\cH{{\mathcal{H}}}
\def\cV{{\mathcal{V}}}
\def\cU{{\mathcal{U}}}

\def\bra#1{\langle #1 |}
\def\ket#1{| #1 \rangle}

\begin{document}
\title*{Three different manifestations \protect\newline of the quantum Zeno effect}
\toctitle{Three different manifestations
\protect\newline of the quantum Zeno effect}
% allows explicit linebreak for the table of content
%
%
\titlerunning{Quantum Zeno Effect}
% allows abbreviation of title, if the full title is too long
% to fit in the running head
%
\author{Paolo Facchi%\inst{1}
\and Saverio Pascazio%\inst{1}
}
\authorrunning{Paolo Facchi and Saverio Pascazio}
% if there are more than two authors,
% please abbreviate author list for running head
%
%
\institute{Dipartimento di Fisica, Universit\`a di Bari, I-70126 Bari, Italy
\\
and Istituto Nazionale di Fisica Nucleare, Sezione di Bari,
I-70126 Bari, Italy
%\and Princeton University, Princeton NJ 08544, USA
}

\maketitle              % typesets the title of the contribution

\begin{abstract}
Three different manifestations of the quantum Zeno effect are
discussed, compared and shown to be physically equivalent. We look
at frequent projective measurements, frequent unitary ``kicks" and
strong continuous coupling. In all these cases, the Hilbert space
of the system splits into invariant ``Zeno" subspaces, among which
any transition is hindered.

\end{abstract}

\section{Introduction}
\label{sec-dpw}
\andy{sec-dpw}

The quantum Zeno phenomenon \cite{Misra,Beskow} is usually viewed
as the effect of repeated projections \cite{von} on a quantum
system. It is, however, a more general phenomenon, that can be
best understood in terms of the dynamical time evolution of
quantum systems and fields \cite{zenoreview}. Indeed, a projection
\textit{\`a la} von Neumann is just a handy way to ``summarize"
the complicated physical processes that take place during a
quantum measurement\index{quantum measurements}. The latter is
performed by an external (macroscopic) apparatus and involves
complicated interactions with the environment. The external system
performing the observation need not be a \textit{bona fide}
detection system, that ``clicks" or is endowed with a pointer. It
is enough that the information on the state of the observed system
be encoded in some external degrees of freedom. Moreover, the
interaction between the system and its environment can be fast (as
compared to any other timescales involved) or slow. In the former
case one says that a ``spectral decomposition" \textit{\`{a} la}
Wigner takes place \cite{Wigner63,PN94} [namely a rapid (and
unitary) physical process that associates different (external)
states to different values of the observable being measured]; in
the latter case one says that a ``continuous" measurement process
occurs \cite{Peres80,Schulman98,zenoreview}.

For instance, a spontaneous emission process is often a very
effective measurement, for it is irreversible and leads to an
entanglement of the state of the system (the emitting atom or
molecule) with the state of the apparatus (the electromagnetic
field). The von Neumann rules arise when one traces away the
photonic state and is left with an incoherent superposition of
atomic states. In the light of these observations, it became clear
in the 80's that the main physical features of the Zeno
effect\index{quantum Zeno effect} would still be apparent if one
would formulate the measurement process in more realistic terms,
introducing a physical apparatus, a Hamiltonian and a suitable
interaction with the measured system, with no explicit use of
projection operators.

More to this, once one has realized that the quantum Zeno effect
(QZE) is a mere consequence of the dynamics and cannot be ascribed
to the ``collapse" of the wave function, one would like to
understand which features of the dynamical process are essential
for observing a QZE. It turns out that not only a \textit{bona
fide} detection scheme, but even
irreversibility\index{irreversibility} is unnecessary. In fact,
\emph{any} interactions -- and not only those that can be
considered as a measurement process of some sort -- that
considerably affect the system (in a sense to be made more precise
in the following) provoke QZE. Therefore, the QZE appears in a
much broader context than its original formulation, whenever a
strong disturbance dominates the time evolution of the quantum
system.

The aim of this article is to discuss three different
manifestations of QZE. We start in Sect.\ \ref{sec-nonselect} with
general (projective) measurements, then extend in Sect.\
\ref{sec-qmaps} the notion of QZE to the case of unitary kicks
\cite{bang} and finally discuss in Sect.\ \ref{sec-contQZE}
(unitary) continuous interactions \cite{theorem}. We show in
Sect.\ \ref{sec-dynssQZE} that, in all these cases, the system is
forced to evolve in a set of orthogonal (``Zeno")
subspaces\index{quantum Zeno subspaces} of the total Hilbert
space. The quantum Zeno subspaces\index{quantum Zeno subspaces}
are completely determined by the disturbance: they are nothing but
its invariant subspaces; in the two latter cases (unitary kicks
and continuous coupling) they are the eigenspaces of the
interaction. An example is considered in Sect.\ \ref{sec-example}
and discussed in Sect.\ \ref{sec-moreex}. We conclude with a rapid
overlook in Sect.\ \ref{sec-concl}.

\section{Projective measurements}
 \label{sec-nonselect}
 \andy{nonselect}

We first consider the case of \textit{bona fide}
measurements\index{quantum measurements}, described by projection
operators \textit{\`{a} la} von Neumann \cite{von}. The
measurements are in general:
\begin{itemize}
\item ``incomplete," in the sense that some outcomes may be lumped
together (for instance because the measuring apparatus has
insufficient resolution); this means that the projection operator
that selects a particular lump is multidimensional (and in this
sense the information gained on the measured observable is
incomplete); \item ``nonselective," in the sense that the
measuring apparatus does not select the different outcomes, but
simply destroys the phase correlations between some states,
provoking the transition from a pure state to a mixture.
\end{itemize}
See, for example, \cite{Schwinger59,Peres98}.

Let us outline the extension \cite{theorem} of Misra and
Sudarshan's theorem \cite{Misra} on the QZE to the case of
incomplete and nonselective measurements. Let Q be a quantum
system, whose states belong to the Hilbert space ${\cal H}$ and
whose evolution is described by the superoperator
\beq
\hat U_t \rho=U(t) \rho U^\dagger(t),\qquad U(t)=\exp(-i H t)
\eeq
where $\rho$ is the density matrix of the system and $H$ a
time-independent lower-bounded Hamiltonian. Let
\beq
\{P_n\}_n, \qquad
P_nP_m=\delta_{mn}P_n,\qquad  \sum_n P_n=1 ,
\eeq
be an orthogonal resolution of the identity and $P_n\cH=\cH_{n}$
the relative subspaces. The associated partition on the total
Hilbert space is
\beq
\label{eq:partition}
\cH=\bigoplus_n \cH_{n}.
\eeq
The nonselective measurement is described by the superoperator
\beq
\label{eq:superP} \hat P \rho=\sum_n P_n \rho P_n
\eeq
and the evolution after $N$ measurements in a time $t$ is governed
by the superoperator
\beq
\label{eq:Nevol}
\hat V^{(N)}_t=\underbrace{\left(\hat P\hat U_{t/N}\right)
\left(\hat P\hat U_{t/N}\right)\cdots \left(\hat P\hat
U_{t/N}\right)}_{N \mathrm{ times}} =\left(\hat P \hat
U_{t/N}\right)^{N} .
\eeq
We perform a first, preparatory measurement, so that the initial
state is
\beq
\label{eq:inrho}
\hat P \rho_0=\sum_n P_n \rho_0 P_n.
\eeq
By assuming the time-reversal invariance and the existence of the
strong limits ($t>0$)
\andy{slims}
\beq
\cV_n (t)=\lim_{N\to\infty} \left[P_n
U\left(\frac{t}{N}\right)\right]^N , \qquad \lim_{t \rightarrow
0^+} \cV_n(t) = P_n , \quad \forall n \
\label{eq:slims}
\eeq
one can show that the operators $\cV_n(t)$ exist for all real $t$
and form a semigroup and that the final state, engendered by the
limiting superoperator
\beq
\label{eq:limseqV}
\hat \cV_t \equiv \lim_{N\to\infty} \hat V^{(N)}_t,
\eeq
is
\andy{rhoZ}
\barr
\rho(t)=\hat \cV_t\rho_0 =\sum_n \cV_n(t) \rho_0 \cV_n^\dagger(t),
\quad \mathrm{with} \quad \sum_n \cV_n^\dagger(t)\cV_n(t)=\sum_n
P_n=\mathbf{1} .\;\; \label{eq:rhoZ}
\earr
The components $\cV_n(t) \rho_0 \cV_n^\dagger(t)$ make up a block
diagonal matrix: the initial density matrix is reduced to a
mixture and any interference between different subspaces $\cH_{n}$
is destroyed (complete decoherence). Moreover,
\andy{probinfu}
\beq
p_n(t) =  \mathrm{Tr} \left[\rho(t)
P_n\right]=\mathrm{Tr}\left[\cV_n(t)\rho_0
\cV_n^\dagger(t)\right]= \mathrm{Tr}\left[\rho_0 P_n\right]=p_n(0)
, \quad \forall n .
\label{eq:probinfu}
\eeq
In words, probability is conserved in each subspace and no
probability ``leakage" between any two subspaces is possible: the
total Hilbert space splits into invariant ``Zeno"
subspaces\index{quantum Zeno subspaces} $\cH_n$ and the different
components of the density matrix evolve independently within each
sector. One can think of the total Hilbert space as the shell of a
tortoise, each invariant subspace being one of the scales. Motion
among different scales is impossible. (See Fig.\
\ref{fig:tortoise1} in the following.) Misra and Sudarshan's
seminal result is reobtained when $p_n(0)=1$ for some $n$, in
(\ref{eq:probinfu}): the initial state is then in one of the
invariant subspaces and the survival probability in that subspace
remains unity.

An important particular case is when the Hamiltonian is bounded
$\| H\|<\infty$. Then each limiting evolution operator $\cV_n(t)$
in (\ref{eq:slims}) is unitary within the subspace $\cH_{n}$ and
has the form
\andy{cVfin}
\beq
\cV_n (t)= \lim_{N\to\infty} [ P_n U(t/N) P_n]^N
= P_n \exp(-i P_n H P_n t)  \label{eq:cVfin} .
\eeq
More generally, if $\cH_{n}\subset D (H)$ (which is trivially
satisfied for a bounded $H$), then the resulting Hamiltonian $P_n
H P_n$ is self-adjoint and $\cV_n(t)$ is unitary in $\cH_{n}$.
When the above condition does not hold, one can always formally
write the limiting evolution in the form (\ref{eq:cVfin}), but has
to define the meaning of $P_n H P_n$ and study the
self-adjointness of the limiting Hamiltonian $P_n H P_n$
\cite{Friedman72,compactregularize}.

In any case, with the necessary precautions on the meaning of
operators and boundary conditions, the Zeno evolution can be
written
\beq
\cV_n (t)= P_n \exp(-i H_{\mathrm{Z}}t )
\eeq
where
\beq
H_{\mathrm{Z}}=\hat P H =\sum_n P_n H P_n
\label{eq:ZenoHam}
\eeq
is the ``Zeno" Hamiltonian.

\section{Unitary kicks\index{unitary ``kicks"}}
\label{sec-qmaps}
\andy{sec-qmaps}

The formulation of the preceding section hinges upon projections
\textit{\`{a} la} von Neumann. Projections are (supposed to be)
\emph{instantaneous} processes, yielding the collapse of the wave
function (an ultimately nonunitary process). However, one can
obtain the QZE without making use of nonunitary evolutions
\cite{PN94}, by exploiting Wigner's idea of spectral decomposition
\cite{Wigner63}. In this section we further elaborate on this
issue, obtaining the QZE by means of a \emph{generic} sequence of
frequent \emph{instantaneous unitary} processes, that need not be
spectral decompositions. We will only give the main results, as
additional details and a complete proof, which is related to von
Neumann's ergodic theorem \cite{ReedSimon}, can be found in
\cite{bang}.

Consider the dynamics of a quantum system Q undergoing $N$
``kicks" $U_{\mathrm{kick}}$ (instantaneous unitary
transformations) in a time interval $t$
\andy{eq:BBevol}
\beq
\label{eq:BBevol}
U_N(t)=\underbrace{\left[U_{\mathrm{kick}}U\left(\frac{t}{N}\right)
\right]\left[U_{\mathrm{kick}}U\left(\frac{t}{N}\right)
\right]\cdots \left[U_{\mathrm{kick}}U\left(\frac{t}{N}\right)
\right]}_{N \mathrm{ times}} .
\eeq
In the large $N$ limit, the dominant contribution is
$U_{\mathrm{kick}}^N$. One therefore considers the sequence of
unitary operators
\andy{eq:sequence}
\barr
\label{eq:sequence}
V_N(t)&=&U_{\mathrm{kick}}^{\dagger N} U_N(t)=
U_{\mathrm{kick}}^{\dagger N}\left[U_{\mathrm{kick}}
U\left(\frac{t}{N}\right) \right]^N
\earr
and its limit
\andy{eq:limseq}
\beq
\label{eq:limseq}
\cU(t) \equiv \lim_{N\to\infty} V_N(t).
\eeq
One can show that
\andy{eq:eqUz}
\beq
\label{eq:eqUz}
\cU(t)= \exp(-i H_{\mathrm{Z}} t),
\eeq
where
\andy{eq:eqHz}
\beq
\label{eq:eqHz}
H_{\mathrm{Z}} = \hat P H = \sum_n P_n H P_n
\eeq
is the Zeno Hamiltonian, $P_n$ being the spectral projections of
$U_{\mathrm{kick}}$
\andy{eq:specdec}
\beq
U_{\mathrm{kick}}=\sum_n e^{-i\lambda_n} P_n . \quad
(e^{-i\lambda_n}\neq e^{-i\lambda_l}, \; \mbox{for} \; n\neq l)
\label{eq:specdec}
\eeq
Incidentally, notice that in this case the map
$H_{\mathrm{Z}}=\hat P H$ is the projection onto the centralizer
(commutant)
\andy{eq:centro}
\beq
\label{eq:centro}
Z(U_{\mathrm{kick}})=\{X|\; [X, U_{\mathrm{kick}}]=0\}.
\eeq
In conclusion
\andy{eq:evolNZ}
\barr
U_N(t)\sim U_{\mathrm{kick}}^N \cU(t)= U_{\mathrm{kick}}^N \exp(-i
H_{\mathrm{Z}} t) =\exp\left(-i \sum_n  N \lambda_n P_n + P_n H
P_n t \right).
\nonumber \\
\label{eq:evolNZ}
\earr
The unitary evolution (\ref{eq:BBevol}) yields therefore a Zeno
effect\index{quantum Zeno effect} and a partition of the Hilbert
space into Zeno subspaces\index{quantum Zeno subspaces}, like in
the case of repeated projective measurements discussed in Sect.\
\ref{sec-nonselect}. The appearance of the Zeno subspaces is a
direct consequence of the wildly oscillating phases between
different eigenspaces of the kick and hinges upon a mean ergodic
theorem. This is equivalent to a procedure of randomization of the
phases. We will see a similar situation in the next section, in
terms of a strong continuous coupling. Notice that if a projection
is viewed as a shorthand notation for a spectral decomposition
\cite{Wigner63,PN94}, the above dynamical scheme includes, for all
practical purposes, the usual formulation of the quantum Zeno
effect in terms of projection operators.

It is superfluous to stress the analogy of the approach outlined
in this section with the seminal papers on quantum maps and
quantum chaos \cite{qch}. In this context, see \cite{varieidee}.

\section{Continuous coupling}
 \label{sec-contQZE}
 \andy{contQZE}

The formulation of the preceding sections hinges upon
instantaneous processes, that can be unitary or nonunitary.
However, as explained in the introduction, the basic features of
the QZE can be obtained by making use of a continuous coupling,
when the external system takes a sort of steady ``gaze" at the
system of interest. The mathematical formulation of this idea is
contained in a theorem \cite{thesis,theorem} on the (large-$K$)
dynamical evolution governed by a \emph{generic} Hamiltonian of
the type
\andy{HKcoup}
\beq
H_K= H + K H_{\mathrm{c}} ,
 \label{eq:HKcoup}
\eeq
which again need not describe a \textit{bona fide} measurement
process. $H$ is the Hamiltonian of the quantum system investigated
and $H_{\mathrm{c}}$ can be viewed as an ``additional" interaction
Hamiltonian performing the ``measurement." $K$ is a coupling
constant.

Consider the time evolution operator
\barr
U_{K}(t) = \exp(-iH_K t) . \label{eq:measinter}
\earr
In the ``infinitely strong measurement" (``infinitely quick
detector") limit $K\to\infty$, the dominant contribution is
$\exp(-i K H_{\mathrm{c}}t)$. One therefore considers the limiting
evolution operator
\beq
\label{eq:limevol} \cU(t)=\lim_{K\to\infty}\exp(i K
H_{\mathrm{c}}t)\,U_{K}(t),
\eeq
that can be shown to have the form
\beq
\label{eq:theorem} \cU(t)=\exp(-i H_{\mathrm{Z}} t),
\eeq
where
\beq
H_{\mathrm{Z}}=\hat P H =\sum_n P_n H P_n \label{eq:diagsys}
\eeq
is the Zeno Hamiltonian [projection of the system Hamiltonian $H$
onto the centralizer $Z(H_{\mathrm{c}})$], $P_n$ being the
eigenprojection of $H_{\mathrm{c}}$ belonging to the eigenvalue
$\eta_n$
\beq
\label{eq:diagevol}
H_{\mathrm{c}} = \sum_n \eta_n P_n, \qquad (\eta_n\neq\eta_m,
\quad \mbox{for} \; n\neq m) \ .
\eeq
This is formally identical to (\ref{eq:ZenoHam}) and
(\ref{eq:eqHz}). In conclusion, the limiting evolution operator is
\andy{eq:measinterbis}
\barr
U_K(t)\sim\exp(-i K H_{\mathrm{c}}t)\,\cU(t) = \exp\left(-i\sum_n
K  t\eta_n P_n + P_n H P_n t\right) ,
\label{eq:measinterbis}
\earr
whose block-diagonal structure is explicit. Compare with
(\ref{eq:evolNZ}). The above statements can be proved by making
use of the adiabatic theorem \cite{Messiah61}. It is also worth
emphasizing interesting links with the quantum evolution in the
strong coupling limit
\cite{Frasca}.

The idea of formulating the Zeno effect in terms of a ``continuous
coupling" to an external apparatus has often appeared in the
literature of the last two decades \cite{Peres80,varicont}.
However, the first quantitative estimate of the link with the
formulation in terms of projective measurements is rather recent
\cite{Schulman98} (see also \cite{zenoreview}).

\section{Dynamical superselection rules\index{superselection rules}}
 \label{sec-dynssQZE}
 \andy{dynssQZE}

\begin{figure}[t]
\begin{center}
\includegraphics[height=6.5cm]{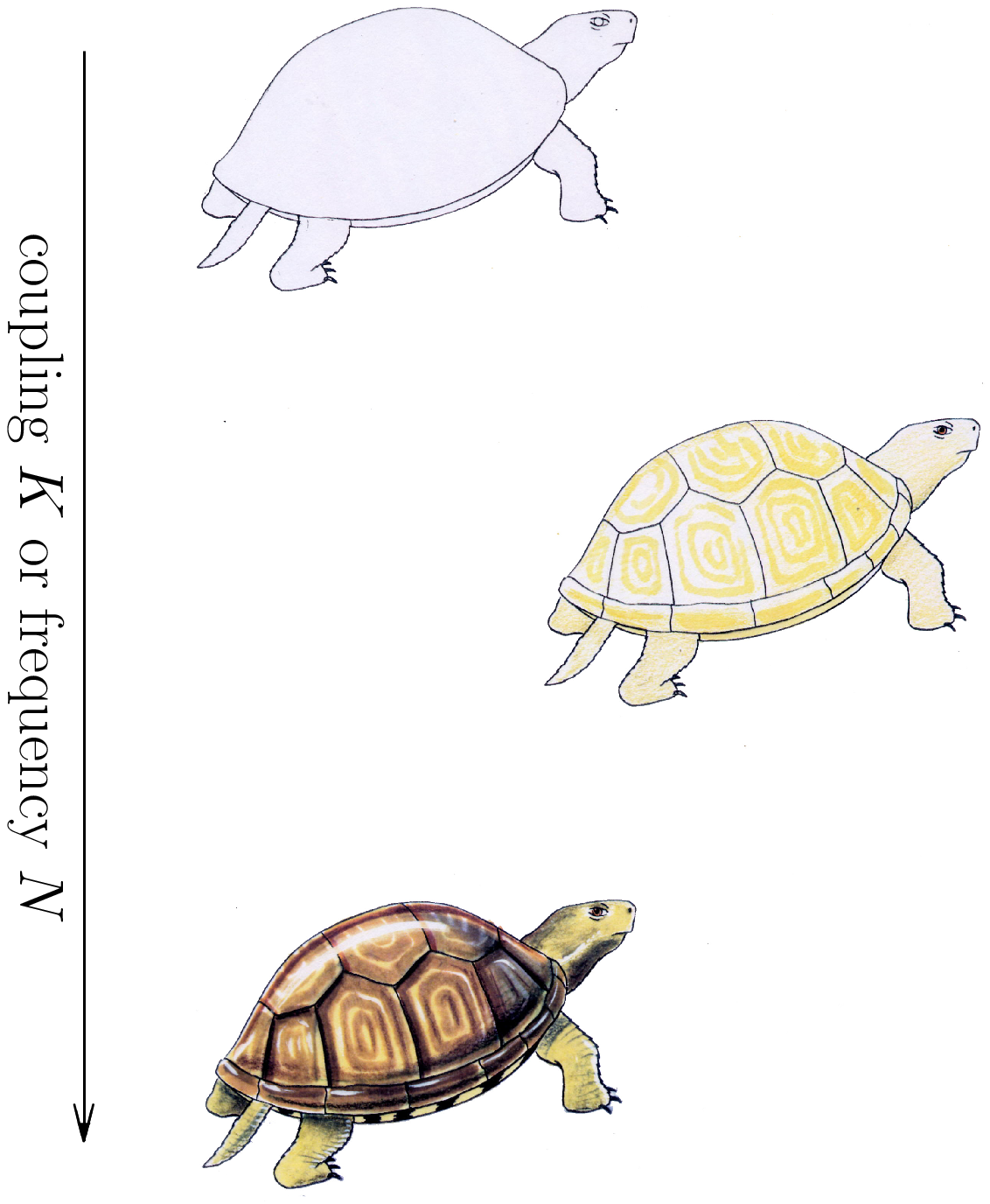}
\end{center}
\caption{\label{fig:tortoise1} The Hilbert space of the system: a
dynamical superselection rule\index{superselection rules} appears
as the number of measurements $N$ or the number of kicks $N$ or
the coupling $K$ to the apparatus are increased}
\end{figure}

Let us briefly consider the main physical implications. In the
$N\to \infty$ ($K\to\infty$) limit the time evolution operator
$\cU(t)$ becomes diagonal with respect to $U_{\mathrm{kick}}$ or
$H_{\mathrm{c}}$,
\beq
[\cU(t), U_{\mathrm{kick}}]=0,\qquad [\cU(t), H_{\mathrm{c}}]=0,
\eeq
a superselection rule arises and the total Hilbert space is split
into subspaces $\cH_{n}$ which are invariant under the evolution.
The dynamics within each Zeno subspace $\cH_{n}$ is essentially
governed by the diagonal part $H_{\mathrm{Z}}P_n =P_n H P_n$ of
the system Hamiltonian $H$ [the remaining part of the evolution
consisting in a sector-dependent phase] and the probability to
find the system in each $\cH_{n}$
\barr
p_n(t)&=&\mathrm{Tr} \left[ \rho(t) P_n \right]= \mathrm{Tr}
\left[\cU(t)\rho_0\cU^\dagger(t) P_n\right] =\mathrm{Tr}
\left[\cU(t)\rho_0 P_n\cU^\dagger(t)\right]
\nonumber\\
&=& \mathrm{Tr} \left[ \rho_0 P_n \right]=p_n(0)
\label{eq:pntpn0}
\earr
is constant. As a consequence, if the initial state is an
incoherent superposition of the form (\ref{eq:inrho}), then each
component will evolve separately, according to
\beq
\rho(t)=\cU(t)\rho_0\cU^\dagger(t)= \sum_n \cV_n(t) \rho_0
\cV_n^\dagger(t),
\label{eq:cUcV}
\eeq
with $\cV_n(t)= P_n \exp(-i P_n H P_n t)$, which is exactly the
same result (\ref{eq:rhoZ})-(\ref{eq:cVfin}) found in the case of
projective measurements. In Fig.\ \ref{fig:tortoise1} we
endeavored to give a pictorial representation of the decomposition
of the Hilbert space in the three cases discussed (projective
measurements, kicks and continuous coupling).

Notice, however, that there is one important difference between
the nonunitary evolution discussed in Sect.\ \ref{sec-nonselect}
and the dynamical evolution discussed in Sects.\
\ref{sec-qmaps}-\ref{sec-contQZE}: indeed, if the initial state
$\rho_0$ contains coherent terms between any two Zeno subspaces
$\cH_{n}$ and $\cH_{m}$, $P_n \rho_0 P_m\neq 0$, these vanish
after the first projection (\ref{eq:rhoZ}) in Sect.\
\ref{sec-nonselect}: $P_n \rho(0^+) P_m = 0$ [the state becomes an
incoherent superposition $\rho(0^+)\neq\rho_0$, whence
$\mathrm{Tr} \rho(0^+)^2 < \mathrm{Tr} \rho_0^2$]. On the other
hand, such terms are preserved by dynamical (unitary) evolution
analyzed in Sects.\ \ref{sec-qmaps}-\ref{sec-contQZE}, and do not
vanish, even though they wildly oscillate. For example, consider
the initial state
\beq
\rho_0= (P_n+P_m)\rho_0 (P_n+P_m), \qquad P_n \rho_0 P_m \neq 0.
\eeq
By (\ref{eq:evolNZ}) and (\ref{eq:measinterbis}) it evolves into
\barr
\rho(t)&=&\cV_n (t) \rho_0 \cV^\dagger_n(t)+ \cV_m (t) \rho_0
\cV^\dagger_m(t)
\nonumber\\
& & + \; e^{-i N (\lambda_n-\lambda_m)} \cV_n (t) \rho_0
\cV^\dagger_m(t) +e^{i N (\lambda_n-\lambda_m)} \cV_m (t) \rho_0
\cV^\dagger_n(t) ,
\\
\rho(t)&=&\cV_n (t) \rho_0 \cV^\dagger_n(t)+ \cV_m (t) \rho_0
\cV^\dagger_m(t)
\nonumber\\
& & + \; e^{-i K (\eta_n-\eta_m) t} \cV_n (t) \rho_0
\cV^\dagger_m(t) +e^{i K (\eta_n-\eta_m) t} \cV_m (t) \rho_0
\cV^\dagger_n(t) ,
\earr
respectively, at variance with (\ref{eq:rhoZ}) and
(\ref{eq:cUcV}), respectively. Therefore $\mathrm{Tr} \rho(t)^2 =
\mathrm{Tr} \rho_0^2$ for any $t$ and the Zeno dynamics is unitary
in the \textit{whole} Hilbert space $\cH$. We notice that these
coherent terms become unobservable in the large-$N$ or large-$K$
limit, as a consequence of the Riemann-Lebesgue theorem (applied
to any observable that ``connects" different sectors and whose
time resolution is finite). This interesting aspect is reminiscent
of some results on ``classical" observables \cite{Jauch},
semiclassical limit \cite{Berry} and quantum
measurement\index{quantum measurements} theory
\cite{Araki,Schwinger59}.

It is worth noticing that the superselection
rules\index{superselection rules} discussed here are \textit{de
facto} equivalent to the celebrated ``W$^3$" ones \cite{WWW}, but
turn out to be a mere consequence of the Zeno dynamics. For a
related discussion, but in a different context, see
\cite{Giulini}.

\section{An example}
 \label{sec-example}
 \andy{example}

One of the main potential applications of the quantum Zeno
subspaces\index{quantum Zeno subspaces} concerns the possibility
of ``freezing" the loss of quantum mechanical coherence and
probability leakage due to the interaction of the system of
interest with its environment. Let us therefore look at an
elementary example in the light of the three different
formulations of the Zeno effect\index{quantum Zeno effect}
summarized in Sections \ref{sec-nonselect}-\ref{sec-contQZE}. In
the following, it can be helpful to think of the Zeno subspace
$\cH_1$ as the quantum computation subspace (qubit) that one wants
to protect from decoherence\index{decoherence}.

\begin{figure}[t]
\begin{center}
\includegraphics[height=3cm]{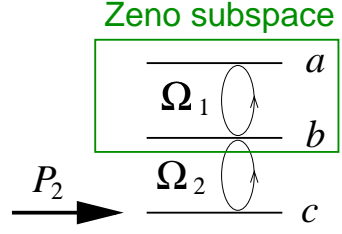}
\end{center}
\caption{Three level system undergoing measurements ($P_1$ not
indicated). We explicitly showed the Zeno subspace $\cH_1$}
\label{fig:fig2}
\end{figure}

Consider a 3-level system in $\cH_{\mathrm{sys}}=\mathbb{C}^3$
\andy{3level}
\beq
\bra{a} = (1,0,0), \quad \bra{b} = (0,1,0), \quad \bra{c} =
(0,0,1)
\label{eq:3level}
\eeq
and the  Hamiltonian
\andy{ham3lev}
\beq
H = \Omega_1 ( \ket{a} \bra{b} + \ket{b} \bra{a}) + \Omega_2 (
\ket{b} \bra{c} + \ket{c} \bra{b}) = \pmatrix{0 & \Omega_1 & 0 \cr
\Omega_1 & 0 & \Omega_2 \cr 0 & \Omega_2 & 0 }.
\label{eq:ham3lev}
\eeq
We perform the (incomplete, nonselective) projective measurements
($P_1+P_2 =\mathbf{1}$)
\andy{proj3}
\beq
P_1 = \ket{a} \bra{a} + \ket{b} \bra{b} = \pmatrix{1 & 0 & 0 \cr 0
& 1 & 0 \cr 0 & 0 & 0 }, \quad P_2 = \ket{c} \bra{c}= \pmatrix{0 &
0 & 0 \cr 0 & 0 & 0 \cr 0 & 0 & 1 },
\label{eq:proj3}
\eeq
yielding the partition (\ref{eq:partition}), with $\mathrm{dim}
\cH_{1}=2$, $\mathrm{dim} \cH_{2}= 1$. The evolution operators
(\ref{eq:cVfin}) read
\andy{evolmeas3}
\barr
\cV_1 &= P_1 \exp(-i P_1 H P_1 t) &= P_1\exp \left[-i\pmatrix{0 &
\Omega_1t & 0 \cr \Omega_1t & 0 & 0 \cr 0 & 0 & 0 } \right]
\nonumber\\
&  &=\pmatrix{\cos\Omega_1t & -i\sin\Omega_1t & 0 \cr
-i\sin\Omega_1t &
\cos\Omega_1t & 0 \cr 0 & 0 & 0 }, \nonumber \\
\cV_2 &= P_2 \exp(-i P_2 H P_2 t) &= P_2= \pmatrix{0 & 0 & 0 \cr 0
& 0 & 0 \cr 0 & 0 & 1 }
\label{eq:evolmeas3}
\earr
and the Zeno Hamiltonian (\ref{eq:ZenoHam}) is
\beq
H_{\mathrm{Z}}= P_1 H P_1 + P_2 H P_2 =\pmatrix{0 & \Omega_1 & 0
\cr \Omega_1 & 0 & 0 \cr 0 & 0 & 0 }.
\eeq
The initial state (\ref{eq:inrho}) evolves according to
(\ref{eq:rhoZ}): in the Zeno limit ($N\to \infty$), the subspaces
$\cH_{1}$ and $\cH_{2}$ decouple. If the coupling $\Omega_2$ is
viewed as a caricature of the loss of quantum mechanical
coherence, the subspace $\cH_{1}$ becomes
``decoherence\index{decoherence} free" \cite{Viola99}. See Fig.\
\ref{fig:fig2}.

In order to understand how unitary kicks yield the Zeno subspaces,
consider the 4-level system in the enlarged Hilbert space
$\cH_{\mathrm{sys}}\oplus \mathrm{span}\{\ket{M}\}$
\andy{levelM}
\beq
\bra{a} = (1,0,0,0), \quad \bra{b} = (0,1,0,0), \quad \bra{c} =
(0,0,1,0), \quad \bra{M} = (0,0,0,1)
\label{eq:levelM}
\eeq
and the Hamiltonian
\andy{ham3l0}
\beq
H = \Omega_1 ( \ket{a} \bra{b} + \ket{b} \bra{a}) + \Omega_2 (
\ket{b} \bra{c} + \ket{c} \bra{b}) = \pmatrix{0 & \Omega_1 & 0 &
0\cr \Omega_1 & 0 & \Omega_2 & 0 \cr 0 & \Omega_2 & 0 & 0 \cr 0 &
0 & 0 & 0 } .
\label{eq:ham3l0}
\eeq
This is the same example as (\ref{eq:3level})-(\ref{eq:ham3lev}),
but we added a fourth level $\ket{M}$. We now couple $\ket{M}$ to
$\ket{c}$ by performing the unitary kicks
\andy{eq:3kicks}
\barr
U_{\mathrm{kick}}&=& e^{-i \lambda_1} P_1 + e^{-i \lambda_2
(\ket{c} \bra{M} + \ket{M} \bra{c})} = \pmatrix{e^{-i\lambda_1} &
0 & 0 & 0\cr 0 & e^{-i\lambda_1} & 0 & 0 \cr 0 & 0 & \cos\lambda_2
& -i\sin\lambda_2 \cr 0 & 0 & -i\sin\lambda_2 & \cos\lambda_2 }
\nonumber\\
&=& \sum_{n=1,\pm} e^{-i\lambda_n} P_n,
\label{eq:3kicks}
\earr
where $\lambda_1 \neq \lambda_2=\lambda_+ = -\lambda_- \neq
\lambda_1$ (otherwise arbitrary), and the subspaces are defined by
\andy{eq:meassub}
\begin{subeqnarray}
 P_1 &=& \ket{a} \bra{a} + \ket{b} \bra{b}= \pmatrix{1 & 0
& 0 & 0\cr 0 & 1 & 0 & 0 \cr 0 & 0 & 0 & 0 \cr 0 & 0 & 0 & 0 } ,
\label{eq:meassuba} \\
 P_\pm &=& \frac{(\ket{c} \pm \ket{M})(\bra{c} \pm
\bra{M})}{2} = \frac{1}{2}\pmatrix{0 & 0 & 0 & 0\cr 0 & 0 & 0 & 0
\cr 0 & 0 & 1 & \pm 1 \cr 0 & 0 & \pm 1 & 1 } .
\label{eq:meassubb}
\end{subeqnarray}
($P_1 + P_- + P_+=\mathbf{1}$.)

\begin{figure}[t]
\begin{center}
\includegraphics[height=3cm]{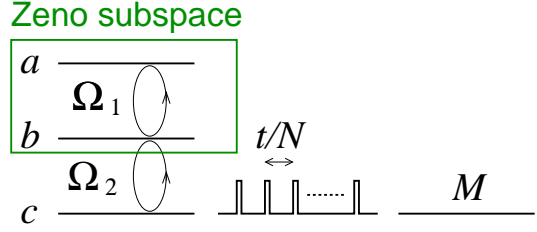}
\end{center}
\caption{Three level system undergoing frequent unitary kicks that
couple one of its levels to an ``external" system $M$.
($\lambda_1=0$.) We explicitly indicated the Zeno subspace
$\cH_1$}
\label{fig:fig3}
\end{figure}

In the Zeno limit ($N\to \infty$) the subspaces $\cH_{1}$,
$\cH_{+}$ and $\cH_{-}$ decouple due to the wildly oscillating
phases O$(N)$. See Fig.\ \ref{fig:fig3}. The Zeno Hamiltonian
(\ref{eq:eqHz}) reads
\andy{eq:eqHz}
\beq
\label{eq:eqHzpp}
H_{\mathrm{Z}} = \sum_n P_n H P_n = \pmatrix{0 & \Omega_1 & 0 & 0
\cr \Omega_1 & 0 & 0 & 0 \cr 0 & 0 & 0 & 0 \cr 0 & 0 & 0 & 0 }
\eeq
and the evolution (\ref{eq:evolNZ}) is
\andy{eq:evolexNZ}
\barr
U_N(t) & \sim & \exp\left(-i \sum_n  N \lambda_n P_n + P_n H P_n t
\right) \nonumber \\
& = & \exp \left[-i\pmatrix{N\lambda_1 & \Omega_1t & 0 & 0\cr
\Omega_1t & N \lambda_1 & 0 & 0 \cr 0 & 0 & 0 & N\lambda_2
\cr 0 & 0 & N\lambda_2 &
0 } \right] .
\label{eq:evolexNZ}
\earr
A natural choice, yielding no phases in the (Zeno) subspace of
interest, is $\lambda_1=0, \lambda_2=1$:
\andy{eq:evolexNZpart}
\barr
U_N(t) &\sim& \exp \left[-i\pmatrix{0 & \Omega_1t & 0 & 0\cr
\Omega_1t & 0 & 0 & 0 \cr 0 & 0 & 0 & N \cr 0 & 0 & N & 0 }
\right] \nonumber\\
&=&\pmatrix{\cos\Omega_1t & -i\sin\Omega_1t & 0 & 0\cr
-i\sin\Omega_1t & \cos\Omega_1t & 0 & 0 \cr 0 & 0 & \cos N &
-i\sin N \cr 0 & 0 & -i\sin N & \cos N } .
\label{eq:evolexNZpart}
\earr

\begin{figure}[t]
\begin{center}
\includegraphics[height=3cm]{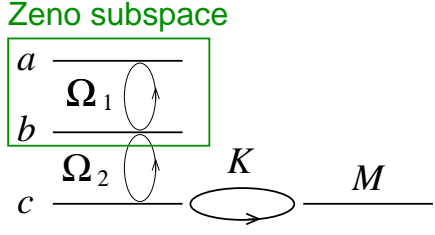}
\end{center}
\caption{Three level system with one of its levels strongly
coupled to an ``external" system $M$.  We explicitly indicated the
Zeno subspace $\cH_1$}
\label{fig:fig4}
\end{figure}

Finally, in order to understand how the scheme involving
continuous measurements works, add to (\ref{eq:ham3l0}) the
Hamiltonian (acting on $\cH_{\mathrm{sys}}\oplus
\mathrm{span}\{\ket{M}\}$)
\andy{hmeas}
\beq
KH_{\mathrm{c}} = K( \ket{c} \bra{M} + \ket{M} \bra{c}) =
\pmatrix{0 & 0 & 0 & 0\cr 0 & 0 & 0 & 0 \cr 0 & 0 & 0 & K \cr 0 &
0 & K & 0 } = K\left(P_{+}-P_{-}\right) \; ,
\label{eq:hmeas}
\eeq
where $P_\pm$ are the same as in (\ref{eq:meassubb}). The fourth
level $\ket{M}$ is now ``continuously" coupled to level $\ket{c}$,
$K \in \mathbb{R}$ being the strength of the coupling
\cite{Militello01}. As $K$ is increased, level $\ket{M}$ performs
a better ``continuous observation" of $\ket{c}$, yielding the Zeno
subspaces. The eigenprojections of $H_{\mathrm{c}}$ [see
(\ref{eq:diagevol})]
\beq
H_{\mathrm{c}}= \eta_1 P_1+ \eta_{-} P_{-} + \eta_{+} P_{+}
\eeq
are again (\ref{eq:meassuba})-(\ref{eq:meassubb}), with $\eta_1=0,
\eta_\pm = \pm 1$. Once again, in the Zeno limit ($K\to \infty$)
the subspaces $\cH_{1}$, $\cH_{+}$ and $\cH_{-}$ decouple due to
the wildly oscillating phases O$(K)$. See Fig.\ \ref{fig:fig4}.
The Zeno Hamiltonian $H_{\mathrm{Z}}$ is given by
(\ref{eq:diagsys}) and turns out to be identical to
(\ref{eq:eqHzpp}), while the evolution (\ref{eq:measinterbis})
explicitly reads
\andy{eq:evolcontZ}
\barr
U_K(t) &\sim & \exp\left(-i \sum_n  Kt \eta_n P_n + P_n H P_n t
\right) \nonumber \\
& = & \exp \left[-i\pmatrix{0 & \Omega_1t & 0 & 0\cr
\Omega_1t & 0 & 0 & 0 \cr 0 & 0 & 0 & Kt \cr 0 & 0 & Kt &
0 } \right]
\nonumber\\
&=&\pmatrix{\cos\Omega_1t & -i\sin\Omega_1t & 0 & 0\cr
-i\sin\Omega_1t & \cos\Omega_1t & 0 & 0 \cr 0 & 0 & \cos Kt &
-i\sin Kt \cr 0 & 0 & -i\sin Kt & \cos Kt }.
\label{eq:evolcontZ}
\earr
[Compare with (\ref{eq:evolexNZpart}): $Kt$ plays the role of
$N$.] It is interesting to notice that with our choice $\eta_1=0$
there are no spurious phases in the (Zeno) subspace of interest.

\section{More examples}
 \label{sec-moreex}
 \andy{moreex}

\subsection{Simplified scheme}

Level $\ket{M}$ and the enlarged Hilbert space
$\cH_{\mathrm{sys}}\oplus \mathrm{span}\{\ket{M}\}$ are not
necessary for our discussion. They were introduced in the example
of the previous section only for the sake of clarity. Consider,
instead of (\ref{eq:3kicks}), the kicks in the original Hilbert
space $\cH_{\mathrm{sys}}=\mathbb{C}^3$
\andy{Ualtern}
\beq
U_{\mathrm{kick}}' =  e^{-i \lambda_1}P_1 + e^{-i \lambda_2}P_2,
\quad (e^{-i \lambda_1} \neq e^{-i \lambda_2})
\label{eq:Ualtern}
\eeq
performed on the \emph{three-level} system
(\ref{eq:3level})-(\ref{eq:proj3}). A straightforward analysis
yields, in the Zeno limit $N \to \infty$, the decomposition
$\cH_{1}\oplus\cH_{2}$. In addition, if $\lambda_1=0,
\lambda_2=1$, then
\andy{Ualternpart}
\beq
U_{\mathrm{kick}}' =P_1+e^{-i }P_2= e^{-iP_2}=e^{-i\ket{c}
\bra{c}}
\label{eq:Ualternpart}
\eeq
and no spurious phases appear.

 Analogously, consider instead of
(\ref{eq:hmeas}), the Hamiltonian in $\cH_{\mathrm{sys}}$
\andy{Haltern}
\beq
H_{\mathrm{c}}' = \eta_1P_1 + \eta_2 P_2  \quad (\eta_1\neq\eta_2)
\label{eq:Haltern}
\eeq
added to (\ref{eq:ham3lev}) in the three-level scheme
(\ref{eq:3level})-(\ref{eq:proj3}). One obtains again, in the $N
\to \infty$ limit, the decomposition $\cH_{1}\oplus\cH_{2}$.
In addition, if $\eta_1=0$ and $\eta_2=1$, then
\andy{Halternpart}
\beq
H_{\mathrm{c}}' =P_2= \ket{c} \bra{c}
\label{eq:Halternpart}
\eeq
(a very simple situation) and no fast oscillating phases appear.

\subsection{Spontaneous decay in vacuum}

As we explained in the previous section, one of the most
interesting potential applications of the quantum Zeno subspaces
concerns the possibility of freezing
decoherence\index{decoherence}, viewed as loss of phase
correlation and/or probability leakage to the environment. The
model outlined in the previous section is too simple to schematize
a genuine decoherence process. For instance, take
(\ref{eq:ham3l0})+(\ref{eq:hmeas}), exemplified in Fig.\
\ref{fig:fig4}: the continuous coupling $K$ does not freeze the
\emph{decay} of level $\ket{b}$ onto level $\ket{c}$, it simply
hinders the \emph{Rabi transition} $\ket{b}\leftrightarrow
\ket{c}$. A better model would be
\andy{hamqc}
\beq
H_K=H_{\mathrm{decay}}+ K H_{\mathrm{c}}= \pmatrix{0 & \Omega_1 &
0 & 0 \cr \Omega_1 & 0 & \tau_{\mathrm{Z}}^{-1} & 0\cr 0 &
\tau_{\mathrm{Z}}^{-1} & -i 2/ \tau_{\mathrm{Z}}^2 \gamma & K \cr
0 & 0 & K & 0}. \label{eq:hamqc}
\eeq
This describes the spontaneous emission of level $\ket{b}$ into a
(structured) continuum, which in turn is resonantly coupled to a
fourth level $\ket{M}$ \cite{zenoreview}. The quantity $\gamma$
represents the decay rate to the continuum and $\tau_{\mathrm{Z}}
= \Omega_2^{-1}$ is the Zeno time \cite{hydrogen} (convexity of
the initial quadratic region). Notice, incidentally, that the Zeno
time can be consistently defined also for open quantum systems
\cite{BenFlor}.

This case is relevant for quantum computation, if one is
interested in protecting a given subspace ($\cH_{1}$) from
decoherence\index{decoherence}, by inhibiting spontaneous
emission. A somewhat related example is considered in
\cite{Agarwal01}. A proper analysis of this model yields to the
following main conclusions: as expected, when the Rabi frequency
$K$ is increased, the spontaneous emission from level $\ket{b}$
(to be ``protected" from decay/decoherence) is hindered. However,
the real problem are the relevant timescales: in order to get an
effective ``protection" of level $\ket{b}$, one needs $K>
1/\tau_{\mathrm{Z}}$. More to this, if the decaying state
$\ket{b}$ has energy $\omega_b\neq 0$, an inverse Zeno effect
\cite{antiZeno,heraclitus} may take place and the requirement for
obtaining the QZE becomes even more stringent \cite{heraclitus},
yielding $K>1/\tau_{\mathrm{Z}}^2\gamma$. Both these conditions
can be very demanding for a real system subject to dissipation.
For instance, typical values for spontaneous decay in vacuum are
$\gamma\simeq 10^9$s$^{-1}$, $\tau_{\mathrm{Z}}^2\simeq
10^{-29}$s$^2$ and $1/\tau_{\mathrm{Z}}^2\gamma\simeq
10^{20}$s$^{-1}$ \cite{hydrogen}. The situation can be made more
favorable by using cavities. In this context, model
(\ref{eq:hamqc}) yields some insights in the examples analyzed in
\cite{Viola99} and \cite{Beige00}, but we will not further
elaborate on this point here.

We emphasize that the case considered in this subsection is not to
be regarded as a toy model. The numerical figures we have given
are realistic and the Hamiltonian (\ref{eq:hamqc}) is a good
approximation of the decay process at short (for the physical
meaning of ``short", see \cite{zenoreview,heraclitus,Antoniou})
and intermediate times (it is not valid for very large times,
where a power law should appear). Related interesting proposals,
making use of kicks or continuous coupling in cavity QED, can be
found in
\cite{Qcomp}. We stress, once again, that the key issue to address
is that of the physically relevant numerical figures and
timescales.

\subsection{Decohering levels}

The example analyzed in Sect.\ \ref{sec-example} essentially aimed
at clarifying that the dynamics in a given subspace $\cH_{1} =
\mathrm{span}\{\ket{a}, \ket{b}\}$ can be decoupled from the
dynamical evolution in the total space. However, it is worth
stressing that level $\ket{a}$ played no role in our discussion.
As a matter of fact, level $\ket{a}$ was simply introduced for
convenience, as only level $\ket{b}$ was interacting with other
levels/systems. Clearly, in a more general framework, one should
look at the decoherence effects on \emph{all} levels that are
coupled to other systems and/or to the environment.

\section{Outlook}
 \label{sec-concl}
 \andy{concl}

We have endeavored to clarify that the main physical features of
the quantum Zeno effect are not peculiar to a quantum measurement
process, but can be framed in a much broader context if one
replaces the projections by a suitable (kicked or continuous)
interaction of the system under investigation (possibly with an
apparatus). However, obviously, whenever the interaction describes
a \textit{bona fide} measurement process performed by a physical
apparatus, one can make use of projection operators
\textit{\`{a} la} von Neumann, if such a description turns out to
be simpler and more economic. This is the principle of Occam's
razor.

A number of important issues have not been discussed in the
present article. Among these, the reach and practical importance
of the experiments on the QZE \cite{expts,Wilkinson,Itanodisc} and
the \emph{physical} meaning of the mathematical expressions
``$N\to \infty$" and ``$K\to\infty$" \cite{heraclitus,zenoreview},
that may involve delicate quantum field theoretical issues
\cite{QFT}. We will not elaborate on this here, but warn the
reader that the expression ``large $N$" and ``large $K$" should
not be taken lightheartedly, as they are directly related to the
physically relevant timescales characterizing the evolution. This
is the key issue to address, in view of possible applications.

%INDEX%%%%%%%%%%%%%%%%%%%%%%%%%%%%%%%%%%%%%%%%%%%%%%%%%%%%%%%%%%%%%%%
% Please check with the editor of your book whether he plans to
% include a "mutual" subject index - if so, please code your entries
% in the standard syntax. For your own purposes you may print your
% "personal" index by using the following commands:
%
%\clearpage
%\addcontentsline{toc}{section}{Index}
%\flushbottom
%\printindex
%%%%%%%%%%%%%%%%%%%%%%%%%%%%%%%%%%%%%%%%%%%%%%%%%%%%%%%%%%%%%%%%%%%%%

\end{document}